\documentclass[preprint,amsmath,noshowkeys,noshowpacs,amssymb, nofootinbib]{revtex4}
\usepackage{graphicx, subfigure}
\usepackage{natbib}
\graphicspath{ {./} }
\usepackage{amssymb, amsmath,amssymb,amsfonts}
\usepackage{amsthm,mathrsfs,amsopn}
\usepackage{dcolumn}
\usepackage{bm}
\usepackage{color}
\usepackage[utf8]{inputenc}
\usepackage{mathtools}
\theoremstyle{plain} %% This is the default
\begin{document}

\title{Multiplicative noise induced bistability and stochastic resonance}

\author{Giuliano Migliorini$^{1,2}$, Duccio Fanelli$^{1}$
 \vspace*{.25cm}}

\affiliation{$^1$ Dipartimento di Fisica e Astronomia, Universit\`a di Firenze, INFN and CSDC, Via Sansone 1, 50019 Sesto Fiorentino, Firenze, Italy}
%\affiliation{$^2$Universit\'e de Orl\'eans,
%Ch\^{a}teau de la Source, Orl\'eans Cedex 45071, France} 
\affiliation{$^2$ Centre de Biophysique Moléculaire, CNRS, Avenue de la Recherche Scientifique, 45071 Orléans, France}

\begin{abstract}
Stochastic resonance is a well established phenomenon, which proves relevant for a wide range of applications, of broad trans-disciplinary breath. Consider a one dimensional bistable stochastic system, characterized by a deterministic double well potential and shaken by an additive noise source. When subject to an external periodic drive, and for a proper choice of the noise strength, the system swings regularly between the two existing deterministic fixed points, with just one switch for each oscillation of the imposed forcing term. This resonant condition can be exploited to unravel weak periodic signals, otherwise inaccessible to conventional detectors. Here, we will set to revisit the stochastic resonance concept by operating in a modified framework where bistability is induced by the nonlinear nature of the multiplicative noise. A candidate model is in particular introduced which fulfils the above requirements while allowing for analytical progress to be made.  Working with reference to this case study, we elaborate on the conditions for the onset of the generalized stochastic resonance mechanism. As a byproduct of the analysis, a novel resonant regime is also identified which displays no lower bound for the frequencies that can be resolved, at variance with the traditional setting.
\end{abstract} 

\maketitle

\section{Introduction}

Noise is in general perceived as a rowdy source of disturbance, which may disruptively alter the inspected dynamics. As opposed to this simplistic view, noise - be it exogenous or endogenous - holds the potential to trigger a vast plethora of non-trivial responses, ranging from seemingly regular cycles \cite{MKNe:Art, PhysRevE.79.036112, ZANKOC2017504} to patterned phases \cite{PhysRevE.81.056110, PhysRevE.81.046215, VaEp:Art}, which are instead lacking under the corresponding deterministic scenario \cite{ZaFaGiLi:Art, FaGiLiZaZa:Art}. In this respect, noise contributes to cooperatively shape the emergent dynamics by enhancing the degree of inherent plasticity for a given system under scrutiny beyond the limited horizon of determinism \cite{ToKa:Art, BiDyMK:Art}. A noticeable example of the constructive endowment ascribed to noise, is constituted by the phenomenon of the so called stochastic resonance \cite{BeSuVu:Art}. Firstly introduced in the context of climatological studies \cite{BePaSuVu82:Art}, the concept of stochastic resonance has rapidly gained widespread importance as a viable strategy to untwist signals too weak to be detected by a conventional sensor \cite{GaHaJuMa:Rev}. More specifically, stochastic resonance occurs in bistable systems, when a small periodic force is applied together with a large wide band noisy perturbation. The interplay of deterministic and stochastic drives, makes in fact the system to swap between the two stable states. When the noise is small, very few switches occur at random, with no characteristic periodicity. Conversely, when the noise is set to be strong, a large number of switches take place for every period of the injected signal to be detected. Remarkably enough, in between these two limiting conditions, an optimal value of the noise exists that cooperatively concurs with the external forcing so as to yield - almost exactly - one switch per period. The onset of this favourable situation can be quantitatively anticipated by comparing two distinct timescales: the period of the imposed regular oscillations - the deterministic time scale - and the Kramers transition rate - the inverse of the stochastic time scale. To sum up, the mechanism of the stochastic resonance builds on the efficient mix of three different ingredients: (i) a double well deterministic potential, to support the bistable dynamics; (ii) an additive noise, namely a random signal with equal intensity at different frequencies or constant power spectral density; (iii) a weak sinusoidal perturbation to be eventually resolved \cite{GaHaJuMa:Rev}. Can one relax the above assumptions - notably (i) and (ii) - while still entraining the cycles of successive swapping to the weak periodic signal? Working along these lines it was shown that the stochastic resonance mechanism can manifest in an over-damped linear system, due to dichotomous noise, a two-valued stochastic process, with constant transition rates between the two allowed states \cite{Fu:Art, BeGi:Art}. In this work, we shall operate under a different angle to provide a positive answer to the above raised question, which in turn implies challenging the stochastic resonance mechanism beyond its original descriptive framework. 
In particular, we shall thoroughly study the case of a system subject to a non linear potential with just one minimum, shaken by a multiplicative noisy source. This latter is appositely devised to slow down the dynamics of the system in correspondence of punctual spatial locations which hence materialise as veritable - noise seeded - fixed points. As we shall prove, and for a proper choice of the parameters involved, the system swings regularly between the two fictitious fixed points, in perfect synchrony with the imposed  sinusoidal forcing. As a side observation, we will also identify a different resonant scheme, which displays no lower bound for the frequencies that can be detected (for any given noise intensity), as opposed to what customarily claimed  working within the original stochastic resonance formulation. The next Section is devoted to investigating the conditions that underlie noise triggered bistability, for a system subject to a single well potential.

\section{Bistability induced by multiplicative noise}

Let us consider a one dimensional stochastic homogeneous system, described by the following stochastic differential equation:
\begin{equation}
\label{eq:Sys}
dx=-\frac{dV}{dx}dt+\epsilon b(x)dW  
\end{equation}
with
\begin{itemize}
    \item $\langle dW(t)\rangle = 0$
    \item $\langle dW(t)dW(t')\rangle = \delta \left(t-t'\right)$
\end{itemize}

The function $b(x)$ defines the multiplicative nature of the  stochastic modulation. Moreover, the amplitude factor $\epsilon$ can be tuned to control the strength of the imposed perturbation.   $V(x)$ represents a deterministic potential, that we require attractive towards a single minimum. This is the only stable fixed point of the deterministic dynamics. With no loss of generality we can assume the minimum to be located in the origin. The above assumptions translates in the obvious constraints $\dfrac{dV}{dx}(x)\leq 0$ if $x<0$ and $\dfrac{dV}{dx}(x)\geq 0$ if $x>0$.  

To proceed further we assume $b(x)$ to be symmetric with respect to $x=0$. Furthermore, we require that in $\pm x_{bar}$ the following conditions hold:
\begin{itemize}
\item $b\left(\pm x_{bar}\right)=0$
\item $\frac{db}{dx}\left(-x_{bar}\right)>0$ 
\item $\frac{db}{dx}\left(+x_{bar}\right)<0$. 
\end{itemize}
The above assumptions amount to posit the existence of two effective barriers located in $\pm x_{bar}$. A quite general choice for the function $b(x)$ is therefore:
\begin{equation}
\label{eq:b(x)}
b(x)=\left( x_{bar}^{2}-x^{2}\right)^{\alpha}
\end{equation}
with $\alpha>0$. 

As follows the above choice, fluctuations are magnified close to the origin, where the deterministic fixed point belongs. Conversely,  random perturbations fade progressively away when approaching the zeroes of the function $b(x)$. This could eventually translate in a veritable freezing of the ensuing dynamics at $\pm x_{bar}$, for suitable choices of the selected deterministic drive.  To state it differently, the system under scrutiny could pass rapidly through the deterministic minimum, to be eventually trapped in the vicinity of the two symmetric states $\pm x_{bar}$. These latter act therefore as a veritable fixed point of the stochastic dynamics. The system could hence manifest a genuine bistable behavior triggered by the very same fluctuating component that is ultimately responsible for the onset of the two aforementioned effective attractors. 

To gain further insight into the above scenario,  consider the forward Fokker-Planck equation associated to the investigated stochastic differential equation (\ref{eq:Sys}), as follows the Ito interpretation \footnote{In principle, we could have chosen to operate under the alternative Stratonovich framework, which would have led to a different forward Fokker-Planck equation. Remark, however, that the two above mentioned interpretations are related by a change of variable. So, working under the Ito ansatz will not affect the generality of our conclusions. Numerical simulations will be performed by means of the celebrated Eulero-Maruyama algorithm for internal coherence.}:
\begin{equation}
\frac{\partial p\left(x,t\right)}{\partial t}=-\frac{\partial }{\partial x}\left(a(x)p\left(x,t\right)\right)+\frac{{\epsilon}^2}{2} \frac{{\partial}^{2}} {{\partial x}^ {2}}\left({b^{2}\left(x\right)}p\left(x,t\right)\right)
\end{equation}
where $p(x,t)$ is the time dependent probability density function to find the system in state $x$ at time $t$.
By setting to zero  the time derivative one obtains an ordinary differential equation for the stationary probability $P_s(x)$:
\begin{equation}
\frac{d}{dx}\left(a(x)P_{s}(x)\right)-\frac{{\epsilon}^2}{2}\frac{d^{2}} {{dx}^ {2}}\left({b^{2}\left(x\right)}P_{s}(x)\right)=\frac{d}{dx}J(x)=0
\end{equation}
where $J(x)$ stands for the probability current. To obtain an explicit solution for $P_s(x)$ we must preliminarily focus  on boundary conditions. Since $b(\pm x_{bar})=0$, in $x=\pm x_{bar}$  the behaviour of the system is reduced to a purely deterministic scheme. The choice of an overall attractive potential, around the deterministic stationary solution $x=0$, implies that the system cannot exit the interval $I=(-x_{bar},x_{bar})$. Indeed, it can be shown that this interval represents the asymptotic attractor for the system. To determine the sought distribution $P_s(x)$, we therefore impose reflecting boundary conditions at $\pm x_{bar}$\footnote{Particularly relevant is the setting in which not only $b(\pm x_{bar})=0$, but also $\frac{d}{dx}V(\pm x_{bar})=0$. This choice amounts to assume that the barriers are inflection points of the deterministic potential. This special class of solution (which corresponds to setting $\epsilon'=0$ in the application discussed below) will not be addressed here \cite{Ga:Testo}.}. The stationary current is everywhere equal to zero and, thus, the stationary probability distribution can be cast in the closed form:
\begin{equation}
P_{s}(x)=\frac{1}{N}e^{-\frac{2}{{\epsilon}^{2}} U_{eff}(x)}.
\end{equation}
where $N\equiv\int_{-x_{bar}}^{x_{bar}}dx e^{-\frac{2}{{\epsilon}^{2}} U_{eff}(x)}$ and where we have introduced the effective potential $U_{eff}(x)$, defined as:
\begin{equation}
\begin{gathered}
\label{eq:Ueffgen}
U_{eff}(x)=\int_{x_0}^x {dx'\frac{dU_{eff}}{dx'}}=\\
=\int_{x_0}^x {dx'\frac{1}{b^2(x')}\left[\frac{dV}{dx'}+{\epsilon}^2b(x')\frac{db}{dx'}(x')\right]}
\end{gathered}
\end{equation}
The potential $U_{eff}(x)$ incorporates all the information about the stationary regime. The form of $U_{eff}(x)$ for $x\rightarrow \pm x_{bar}$ reverberates on the profile of $P_s(x)$. It can be straightforwardly shown that $\lim_{x\to {x _{bar}}_-}U_{eff}(x)=+\infty\Rightarrow\lim_{x\to {x _{bar}}_-}P_s(x)=0$, provided  $\alpha>\frac{1}{2}$. This is just the setting that we ought to explore: when the latter condition holds, in fact, $P_s(x)$ possesses one (or more) maximum in the examined interval $I$. The asymptotic dynamics of the system can be anticipated by inspection of the inequality:
\begin{equation}
\label{eq:d/dxUeff(x)}
\frac{d}{dx}U_{eff}(x)=\frac{1}{b^{2}(x)}\left[\frac{dV}{dx}+\frac{{\epsilon}^2}{2}\frac{d}{dx} b^{2}(x) \right]\geq0
\end{equation}
in $I=(-x_{bar},x_{bar})$. This latter allows to quantitatively resolve the antagonistic interplay between two opposing tendencies: on the one side that exerted by the attractive deterministic potential; on the other, the repulsive drive triggered by the nonlinear  multiplicative noise. The existence of multiple minima of $U_{eff}(x)$ is the condition to impose for the onset of a noise driven instability. In the following, we will set to study a specific system which falls in the class of models defined above.

\subsection{A simple toy model for analytical progress to be made.}

As anticipated above, we shall hereafter focus on a specific example to clarify the setting of interest. The system under inspection will obey equation (\ref{eq:Sys}) with the following choice for the deterministic potential $V(x)$:

\begin{equation}
\label{potential}
V(x)=\frac{1}{6}{\left(x^2-1\right)}^3
\end{equation}

\begin{figure}
\includegraphics[width=1\textwidth]{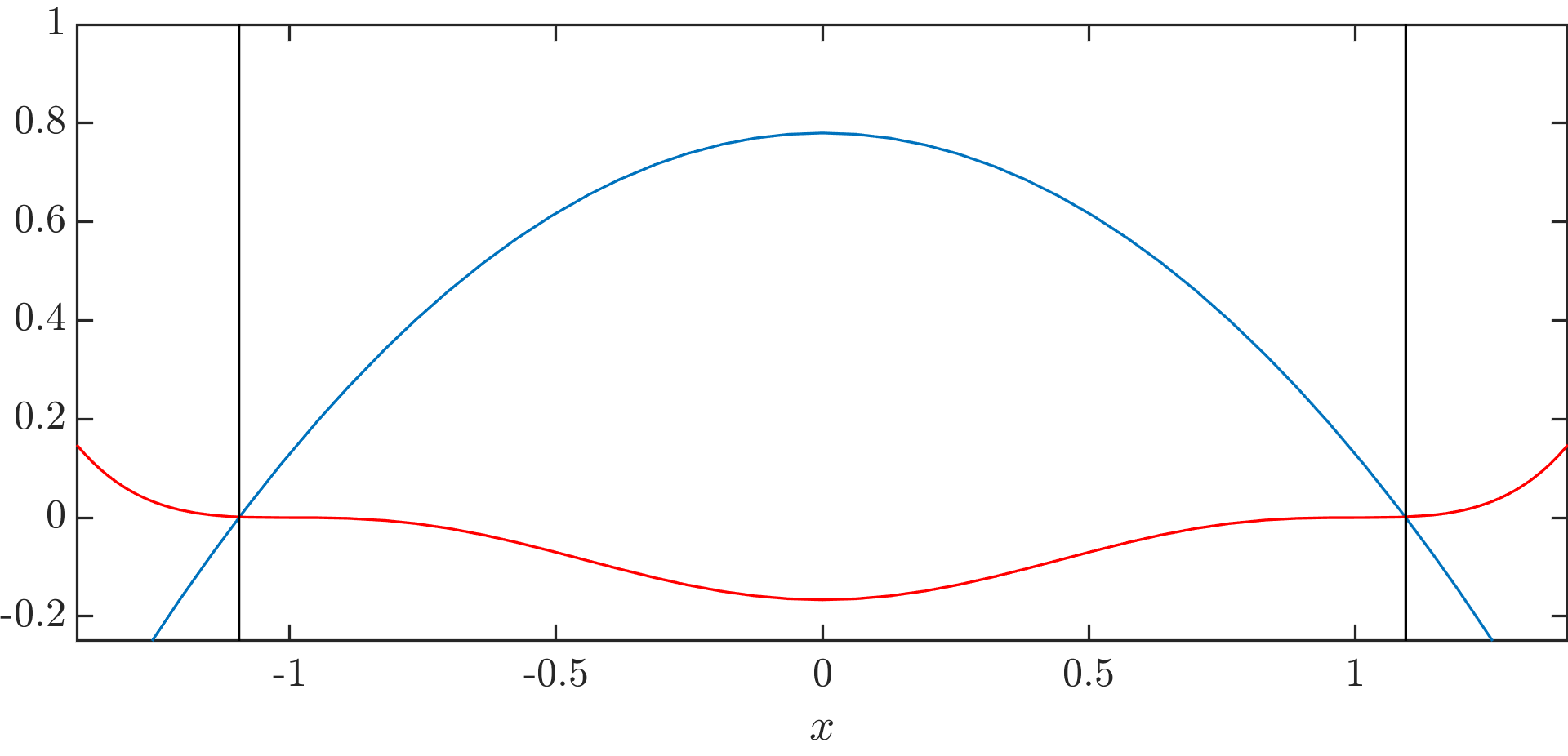}
\caption{The potential $V(x)$ is plotted in red; The noise amplitude factor $\epsilon b(x)=\epsilon (1-\epsilon'-x^2)$ is displayed in blue, for $\epsilon =0.65$ and $\epsilon'=-0.2$. The vertical lines identify the edges of the domain $I$.}
\label{fig:Vb}
\end{figure}

The chosen potential (see profile depicted in red in Figure \ref{fig:Vb}) is attractive towards the origin, the only stable fixed point of the deterministic dynamics and exhibits two symmetric inflection points in $x=\pm 1$. This latter choice yields an effective slowing down of the deterministic dynamics close to the domain of the real axis where the inflections points are located. The introduced potential is however nonlinear and bears paradigmatic relevance. It is in fact the normal form of a pitchfork bifurcation \cite{St:Testo}. Moreover we set:

\begin{equation}
\label{b}
b(x)=1-\epsilon'-x^2 
\end{equation}
to modulate the intensity of the noise as a function of the state variable  $x$. The noise reaches its maximum in $x=0$, namely when the deterministic potential has its minimum. Conversely it is identically equal to zero for $x=\pm x_{bar}=\pm \sqrt{1-\epsilon'}$, if $\epsilon'<1$. With such a choice the noise has a centrifugal effect: it repels the system from the central deterministic stable fixed point,  towards the side barriers. The parameter $\epsilon'$ sets the position of the barriers with respect to the chosen location of the inflection points that characterize the deterministic potential. 

\begin{figure}
\includegraphics[width=1\textwidth]{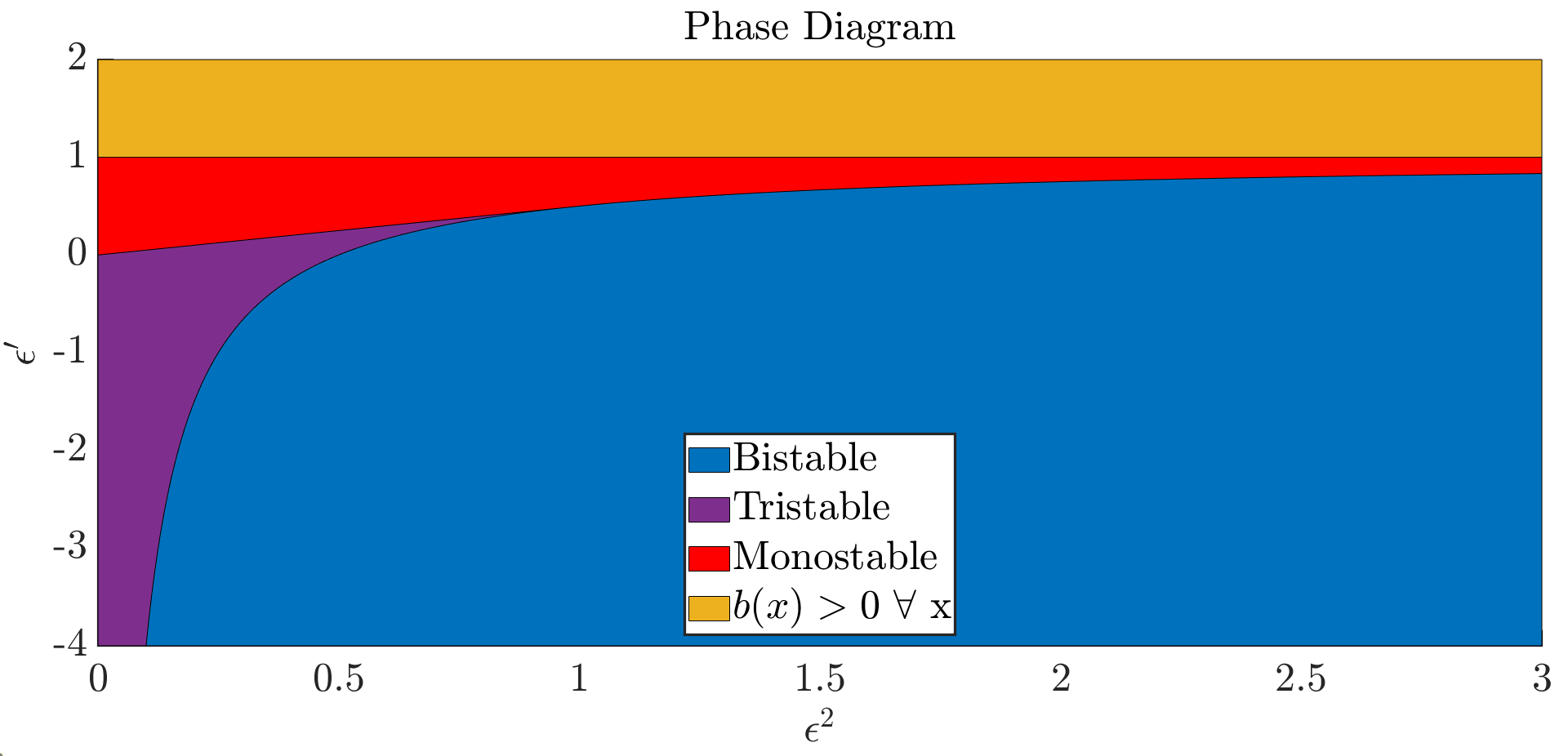}
\caption{Phase diagram for the examined system in the plane ($\epsilon^2, \epsilon'$) The region depicted in orange ($\epsilon'>1$) identifies the portion of the parameters plane where
$b(x)>0$ (the noise term is always active). In the region plotted in red the effective potential exhibits one only minimum, thus yielding an isolated maximum for the  stationary distribution $P_s(x)$.  In the region depicted in purple, eq. (\ref{eq:d/dxUeff(x)}) has five distinct stationary points. The stationary distribution $P_s(x)$ has maxima localized in $x^*=0$ and  $\pm x_{max}$. This latter region exists for $0\leq {\epsilon}^2\leq 1$, and is delimited by the segment $\epsilon'=\frac{{\epsilon}^2}{2}$ and the line $\epsilon'=1-\frac{1}{2{\epsilon}^2}$. This latter curve sets the boundary with the bistable phase (shown in blue). Crossing this line, the point $x^*=0$ loses its stability and becomes a minimum for $P_s(x)$. In this phase, the stationary probability distribution $P_s(x)$ displays two symmetric maxima at finite distance from the origin. }
\label{fig:DiagFase}
\end{figure}

Notice that the system is invariant under the change of variables $x\rightarrow -x$. Following the general reasoning introduced above, we will hereafter prove that system (\ref{eq:Sys}) complemented with   (\ref{potential}) and (\ref{b})
exhibits a noise driven bistable phase.
Under these operating conditions, equation (\ref{eq:Ueffgen}) can be integrated analytically, yielding:
\begin{equation}
\begin{gathered}
\label{eq:UeffSys}
U_{eff}(x)=\frac{{\epsilon'}^2}{2\left(1-\epsilon'\right)}\left[\frac{x^2}{1-\epsilon'-x^2}\right]+\\
-\epsilon'\log\left(\frac{1-\epsilon'-x^2}{1-\epsilon'}\right)+{\epsilon}^2\log(1-\epsilon'-x^2)+\frac{1}{2}x^2
\end{gathered}
\end{equation}
where we assumed $U_{eff}(x=0)={\epsilon}^2\log(1-\epsilon')$ with no loss of generality. Different dynamical behaviours can be identified depending on the chosen parameters $\epsilon$ and $\epsilon'$, as schematically illustrated in figure \ref{fig:DiagFase}, where the 
response of the system is studied by varying $\epsilon'$ and $\epsilon$.
More specifically, the sought bistable regime is predicted to occur for $\epsilon > \frac{1}{2(1-\epsilon')}$, provided $\epsilon'>0$. 
As expected, the bistable phase emerges for sufficiently large  values of $\epsilon$, corresponding to significant noise strengths. This is the setting that we shall thoroughly inspect in the following Section.  In this regime $U_{eff}(x)$ materializes in a double-well potential - analogous to the deterministic double well potential which anticipates a prototypical bistable behavior - and diverges for $x\rightarrow \pm x_{bar}$.
 
\begin{figure}
\includegraphics[width=1\textwidth]{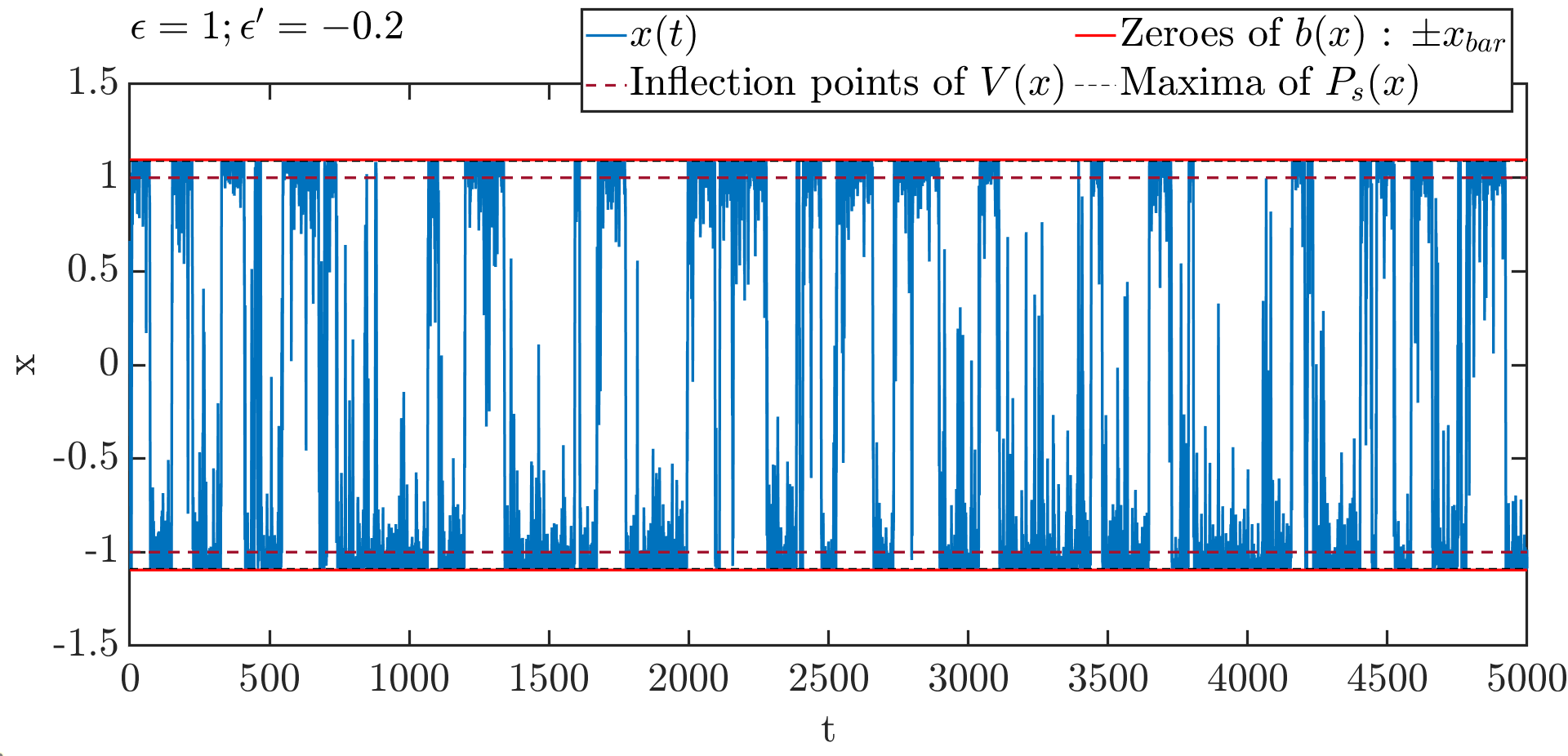}
\caption{Example of bistable dynamics for $\epsilon=1$ and $\epsilon'=-0.2$. A sequence of successive of hops between the maxima of $P_s(x)$ can be observed. For the chosen parameters the maxima of $P_s(x)$ are located at $\pm x_{max}\simeq\pm 1.088$, very close to the barriers, which are positioned in $\pm x_{bar}=\sqrt{1-\epsilon'}\simeq \pm 1.095$. }
\label{fig:bist}
\end{figure}

Focus now on the detailed characterization of the bistable phase. An illustrative example of the dynamics experienced by the stochastic system in this regime can be found in Figure \ref{fig:bist}. 

\begin{figure}
\centering
\includegraphics[width=1\textwidth]{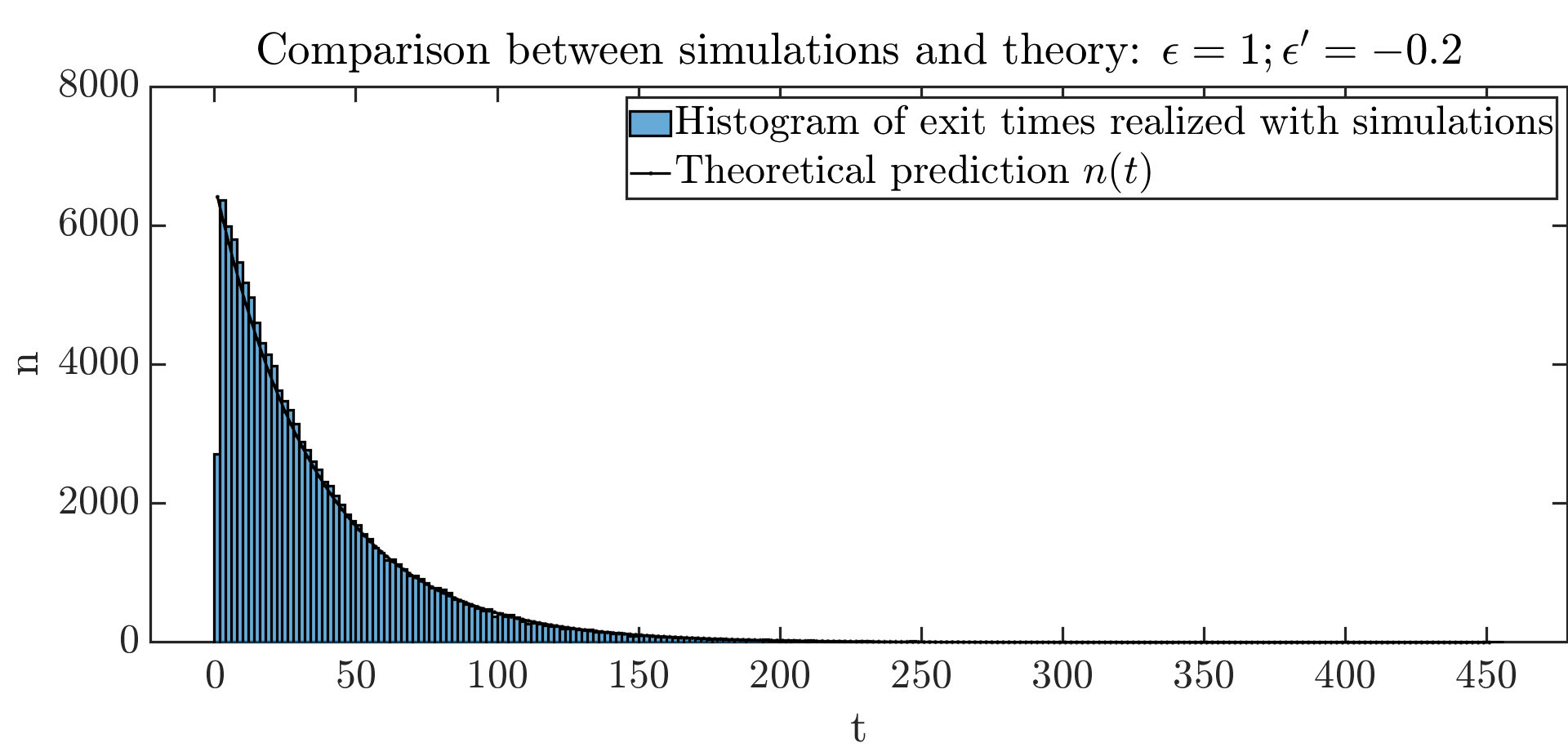}
\caption{The histogram of the recorded exit times from the interval $\left(-x_{bar},0\right)$ is depicted. Here the system is initialized at $x=-x_{max}$, one of the two positions where $P_s(x)$ displays a maximum. The histogram is computed by analysing data from $n = 1.2\cdot 10^5$ independent realizations of the stochastic dynamics. The solid line stands for the theoretical solution discussed in the main body of the paper.}
\label{fig:histteo}
\end{figure}

It is indeed interesting to inspect the distribution of exit times, i.e. the time employed by the stochastic system  to leave one of the maxima ($\pm x_{max}$) of $P_s(x)$ and cross the central barrier to head towards the other maximum. As displayed in Figure \ref{fig:histteo},  the distribution of recorded exit times can be nicely approximated by a decreasing exponential function $f(t)=\frac{1}{\langle \tau \rangle}e^{-\frac{t}{\langle \tau \rangle}}$. This latter is entirely specified by means of a single parameter, the average exit time $\langle \tau \rangle$. Here, $\langle \cdot \rangle$ represents the average over noise (over $n$ realizations for $n\rightarrow \infty$) of the stochastic dynamics. Interestingly, $\langle \tau \rangle$ can be also accessed theoretically by focusing on the interval $(-x_{bar}, 0)$, imposing: (i) a reflecting condition in $-x_{bar}$, and (ii) an absorbing condition in the origin, for the initial condition $x=-x_{max}$. In formulae one gets:
\begin{equation}
\label{eq:tau}
\langle\tau\rangle=\frac{2}{{\epsilon}^2}\int_{-x_{max}}^{0}{dxe^{\frac{2}{\epsilon ^2}U(x)}}\int_{-x_{bar}}^{y}{dx'\frac{1}{b^2(x')}e^{-\frac{2}{\epsilon ^2}U_{eff}(x')}}
\end{equation}
where we have defined:
\begin{equation}
\label{eq:U(x)}
U(x)\equiv\int_{0}^{x}{dx'\frac{1}{b^2(x')}\frac{dV}{dx'}}
\end{equation}

The exponential decaying function with a characteristic time scale $\langle\tau\rangle$ computed as above adheres well to the simulated data, thus confirming the adequacy of the proposed interpretative framework. This observation strengthen the analogy with conventional bistability, and will prove useful in the following to address the generalized analogue of the stochastic resonance. 
It should be also remarked that $\langle \tau^2 \rangle =2{\langle \tau \rangle}^2$. The exponential distribution is an indirect byproduct of the irregular jumps that occur between the two attractive poles, stemming from the multiplicative nature of the noise. Summing up, we have successfully engineered a system which features analogously to those displaying a traditional bistable dynamics,  without imposing a deterministic double well potential.

\section{Stochastic resonance driven by multiplicative noise: the general framework}

The aim of this Section is to discuss the concept of stochastic resonance for the general setting of a noise driven bistable  dynamical unit. We will begin by illustrating the reference framework and then move forward to thoroughly analyse the simple model introduced above. 

To investigate the possible occurrence of stochastic resonance, we account for a periodic forcing as specified in the following: 
\begin{equation}
\label{eq:LangA(t)}
dx=[-\frac{dV}{dx}(x)+ A\cos(\Omega t)]dt+\epsilon b(x)dW
\end{equation}
where $\Omega$ and $A>0$ stands respectively for the frequency and the amplitude of the imposed  forcing. In principle, to elaborate on the impact of the forcing term one should deal with the non autonomous contribution as stemming from the modified deterministic drive. To favour intuition we rather prefer to stick to the approach pioneered in \cite{BeSuVu:Art} and focus on the two limiting settings $ A(t) = \pm \,   A \,  \forall \,  t$. This allows to drop the time dependence and avoid ensuing complications, while accounting for the maximum (signed) perturbations that arise from the periodic modulation. As we will argue, and based on the analysis reported in \cite{BeSuVu:Art}, this working hypothesis allows for a complete characterization of the stochastic resonance phenomenon to be eventually carried out.    

Consider the time-independent stochastic differential equations reported below:
\begin{equation}
\label{eq:LangForzCost}
dx=[-\frac{dV}{dx}(x)\pm A]dt+\epsilon b(x)dW
\end{equation}
for $A<A_c=\frac{dV}{dx}(\pm x_{bar})$. By studying the 
the sign of the perturbed deterministic potential at $\pm x_{bar}$, 
one can readily conclude that the system is indefinitely constrained in the interval $I=(-x_{bar},x_{bar})$, if it gets started inside the very same interval. As for the limiting case $A=0$, one can write down the associated Fokker-Planck stationary equation:

\begin{equation}
\begin{gathered}
\label{eq:dPs/dxA}
\frac{d P_{s}(x)}{dx}=\frac{2}{{\epsilon}^2b^{2}(x)}\left[-\frac{dV}{dx}\pm A-\frac{{\epsilon}^2}{2}\frac{d}{dx} b^{2}(x) \right]P_{s}(x)\equiv \\ \equiv -\frac{2}{{\epsilon}^2}\frac{d}{dx}U_{eff,\pm A}(x)
\end{gathered}
\end{equation}
and impose reflecting boundary conditions in $\pm x_{bar}$. The dynamics of the stochastic system can be anticipated by characterizing the perturbed effective potential $U_{eff,\pm A}(x)$ or, equivalently, the stationary probability density $P_s(x)$, both quantities being self-consistently defined by equation (\ref{eq:dPs/dxA}). Performing a straightforward calculation yields:

\begin{equation}
\begin{gathered}
\label{eq:Ueff+-A}
U_{eff,\pm A}(x)=\int_{x_0}^x {dx'\frac{dU_{eff,\pm A}}{dx'}}= \\
=\int_{x_0}^x {dx'\frac{1}{b^2(x')}\left[\frac{dV}{dx'}\mp A +{\epsilon}^2b(x')\frac{db}{dx'}(x')\right]} 
\end{gathered}
\end{equation}
and
\begin{equation}
P_{s}(x)=\frac{1}{N(\pm A)}e^{-\frac{2}{{\epsilon}^{2}} U_{eff,\pm A}(x)}
\end{equation}
where $N(\pm A)$ is the appropriate normalization constant. 

In the preceding Section we showed that a bistable effective potential $U_{eff}(x)$ could be obtained in absence of an external forcing. For the modified system here considered, and provided the constant forcing $A$ is small enough, we expect to find the same general behavior. By studying $U_{eff,\pm A}(x)$ in $\pm x_{bar}$, it is indeed possible to show that the condition $A<A_c$ makes the potential diverge at the barriers. Furthermore, $U_{eff,\pm A}(x)$ can display two symmetric minima in the interval $I=(-x_{bar},x_{bar})$, a condition - induced by multiplicative noise - which paves the way for the onset of the resonant regime.
To elaborate on this aspect, in the simplified model framework introduced above, it is entirely devoted the following subsection. Before proceeding along these lines, we assume for the sake of complete information that $U_{eff,\pm A}(x)$ gives rise to a double well potential, diverging at the boundaries, which is made asymmetric by the forcing.  A key quantity to gain insight into the conditions that ultimately underlie the inception of the stochastic resonance phenomenon, is represented by the mean exit times $\tau_{\pm A}$. Consider as an initial condition $x(t=0)=-x_{max}$, the maximum of $P_s(x)$ found at $x<0$. Then, the exit times are easily calculated as stipulated by equation (\ref{eq:tau}):
\begin{equation}
\tau_{\pm A}=\frac{2}{{\epsilon}^2}\int_{-x_{max}}^{x_2}{dy e^{\frac{2}{{\epsilon}^2}U_{\pm A}(y)}\int_{-x_{bar}}^y{dx'e^{-\frac{2}{{\epsilon}^2}U_{eff,\pm A}(x')}}}
\end{equation}

Guided by intuition, we can argue that the external forcing breaks the inherent symmetry by preferentially pushing the system towards one of the two maxima as displayed by $P_s(x)$. If the forcing is negative ($-A$), the well of the effective potential $U_{eff,-A}(x)$ at negative $x$ is deeper as compared to its symmetric homologue at positive $x$. The opposite holds for $U_{eff,+A}(x)$ in presence of a positive constant bias $+ A$. With the above-mentioned initial conditions, $\tau_{+ A}$ refers to the mean exit time computed for a forcing which tends to disfavour having the system to evolve in proximity of the maximum of $P_s(x)$ positioned at $x<0$. On the contrary, $\tau_{-A}$ is the mean exit time referred to a setting that pushes the system towards the noise triggered attractor located on the negative portion of the $x$ axis. Focusing on the negative forcing case, clearly, we expect $\tau_{+ A}\leq \tau_{-A}$. Let us now suppose that there is a suitable choice of the parameters, such that the two exit times are significantly different. Following the reasoning in \cite{BeSuVu:Art}, it can be postulated that that the cycles of successive swapping from one well to the other can be entrained to a weak periodic signal whose frequency $\Omega$ matches the condition

\begin{equation}
\label{eq:BeSuVucond}
\langle\tau(+ A)\rangle<<\frac{\pi}{\Omega}\leq\langle\tau(- A)\rangle
\end{equation} 

The better the condition (\ref{eq:BeSuVucond}) is matched, the less the distributions of exit times corresponding to the limiting cases of favourable or opposing external signals overlap. If the periodic forcing fulfills the condition (\ref{eq:BeSuVucond}), it is highly probable that all "fast jumps", represented here by the mean exit time $\langle\tau(+ A)\rangle$, occur in a semi-period of the forcing. On the other hand, the probability that a "slow jump", with average periodicity $\langle\tau(- A)\rangle$ takes place before a semi-period, is low. When subject to a proper periodic forcing, the resulting dynamics appears effectively in-phase with the external signal. This latter qualitative picture, inspired to the seminal work \cite{BeSuVu:Art}, applies also to the relevant setting where the underlying bistability is shaped by the multiplicative and non linear noisy component.

\subsection{Validating the proposed framework with reference with the introduced toy model}

Consider the simple model introduced above and characterized by conditions (\ref{potential}) and (\ref{b}). Following equation (\ref{eq:Ueff+-A}) and imposing,  $U_{eff,\pm A}\left(x=0\right)=U_{eff}\left(x=0\right)=\epsilon^2\log(1-\epsilon')$, as done in equation (\ref{eq:UeffSys}), one can readily obtain an explicit expression for $U_{eff,\pm A}$ given by:
\begin{equation}
\begin{gathered}
\label{eq:UeffpmASist3}
U_{eff,\pm A}\left(x\right)=U_{eff}(x) \mp  \frac{1}{2}\frac{A}{1-\epsilon'}\frac{x}{1-\epsilon'-x^2}+ \\ \mp\frac{1}{4}\frac{A}{(1-\epsilon')^{\frac{3}{2}}}\log{\left(\frac{x+\sqrt{1-\epsilon'}}{\sqrt{1-\epsilon'}-x}\right)}
\end{gathered}
\end{equation}
For $A<A_c$, the introduction of the external forcing breaks the reflection symmetry $x\rightarrow -x$. As remarked earlier,  $\tau_{-A}\geq\tau_{+A}$. Further, numerical simulations suggest that the distribution of exit times obeys a decreasing exponential function, as for the case with $A=0$. More importantly,  parameters $(\epsilon ,\epsilon ')$ can be chosen to have the time scale separation dictated by condition (\ref{eq:BeSuVucond}). This is in particular achieved for values of $\epsilon$, that sit at the frontier between the bistable and tristable phases, inside the region where the bistable dynamics holds (see Figure \ref{fig:DiagFase}). The frequency $\Omega$ characterising the periodic external forcing is chosen to match (\ref{eq:BeSuVucond}) \footnote{More in depth, $\pi/\Omega$ is
an upper bound to the histogram of fast jumps as empirically recorded via numerical simulations of equation \ref{eq:LangForzCost} for the initial condition $x(t=0)=-x_{max}$. Since the distribution of the exit times is well represented by a decreasing exponential, it is not possible to choose $\Omega$ such that $\pi/\Omega$ is strictly smaller than all the entries of the distribution of the slow jumps. Some hops will occur unavoidably out-of-phase.}.
\begin{figure}
\begin{subfigure}
{\includegraphics[width=1\textwidth]{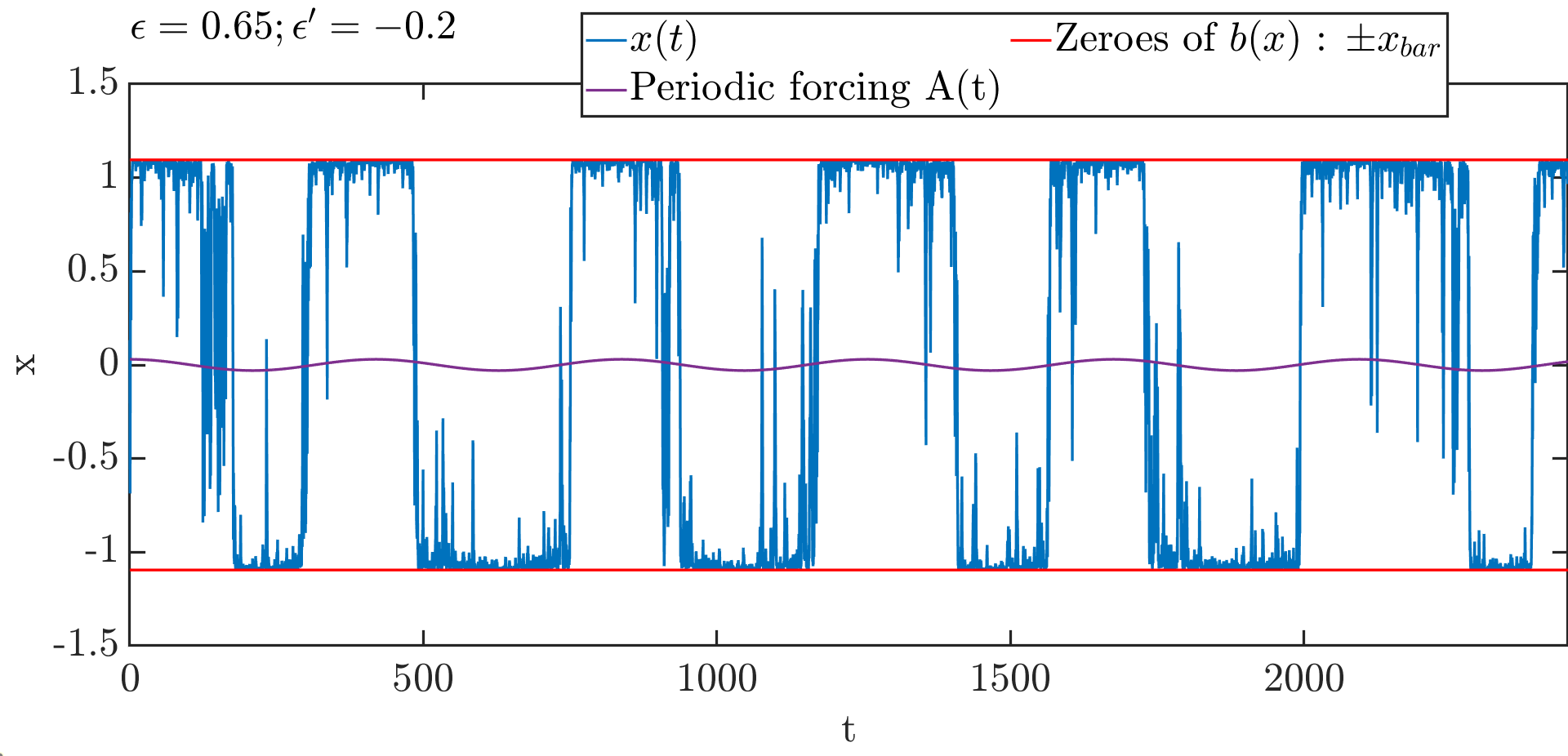}}
\end{subfigure} \quad
\begin{subfigure}
{\includegraphics[width=1\textwidth]{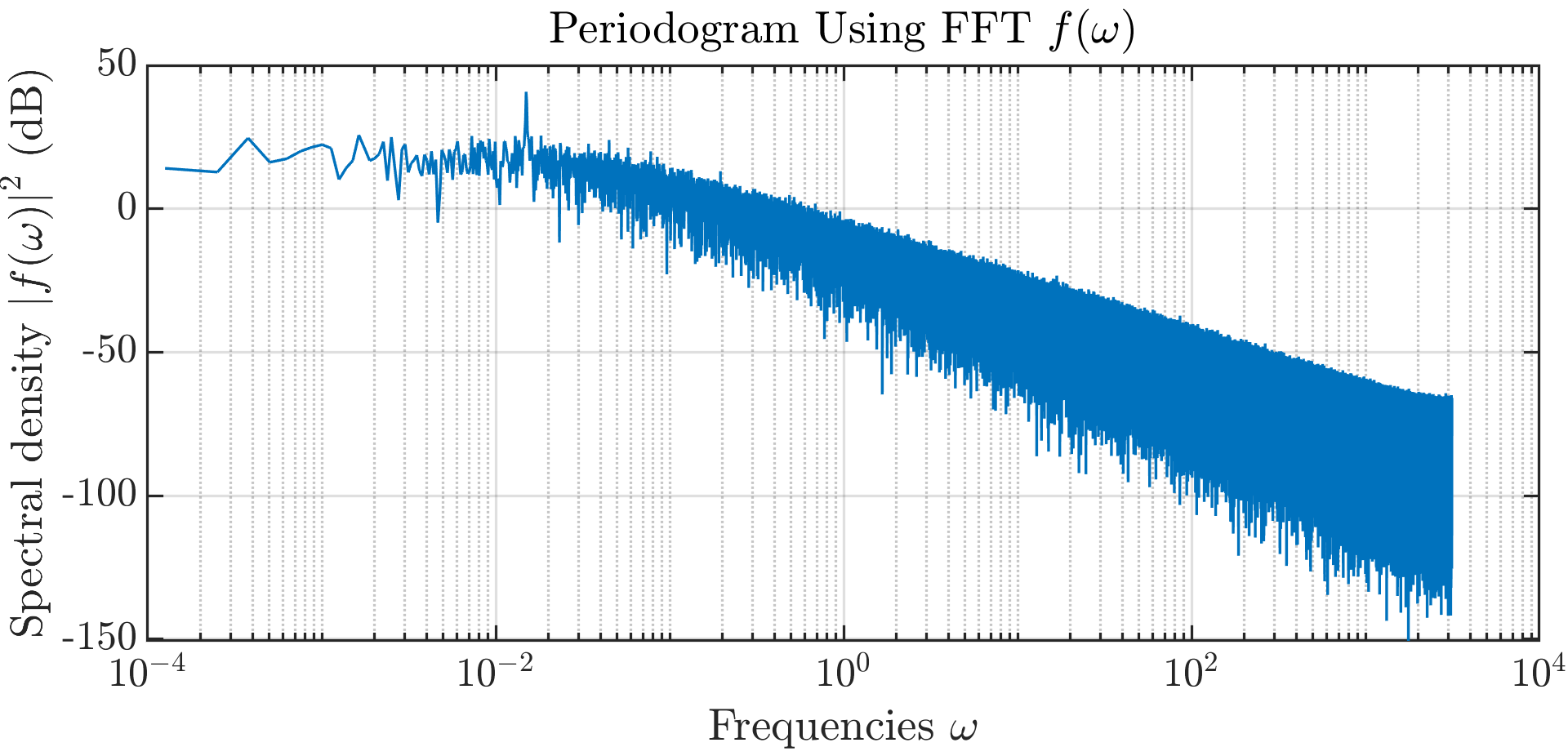}}
\end{subfigure}
\caption{Temporal series and associated power spectrum. The simulated stochastic time series (blue) is synchronized with the imposed periodic modulation (purple), see upper panel. This is also testified by the peak in the power spectrum at the frequency of the forcing $\omega=\Omega$ (lower panel). The response of the system results in the amplification of the input signal. The amplification factor is quantified in $\frac{x_{max}}{A}\simeq\frac{1}{A}\simeq 30$, for this choice of $A$. Here, $\epsilon=0.65,\epsilon'=-0.2,A=0.03,\Omega=0.015$; $\epsilon$ and $\Omega$ are chosen to trigger a significant resonant amplification, while keeping fixed the other parameters.
}
\label{fig:Sys1StocRes1}
\end{figure}

In Figure \ref{fig:Sys1StocRes1} the simulated stochastic time series is depicted in blue and gets synchronized with the imposed periodic modulation (purple). This is also confirmed by the peak at $\omega=\Omega$, as displayed by the reported periodogram. This can be thought as a distinctive feature of the stochastic resonance. The amplification factor is also quantified in the caption of Figure \ref{fig:Sys1StocRes1}. Summing up, we have here shown that a resonant response can manifest in a stochastic system, subject to a trivial single well potential and shaken by a multiplicative noise source, when perturbed by a small periodic external forcing. In the following, and as a remarkable consequence of the setting explored, we will discuss a novel resonant regime which cannot be made available under the usual stochastic resonant paradigm.

\section{A novel regime: gated synchronization}

As stated above, working in the proposed setting it is possible to highlight the emergence of a novel resonant regime, which lacks immediate analogy with the traditional stochastic resonance framework. 
Consider equation (\ref{eq:LangA(t)}) and focus on the simplified limiting setting obtained by replacing the periodic drive with the constant factors $\pm A$. At variance with the above discussion, we  set to study the case $A>A_c=\vert \frac{dV}{dx}(x_{bar})\vert$. 
This condition reflects in a drastic change of the asymptotic behaviour of the system, as we will set to clarify.  Imagine that the system evolves to eventually reach one of the barriers in $\pm x_{bar}$. The dynamics turns then in a purely deterministic scheme, and the system (when $A>A_c$) is pushed outside of the domain $I=(-x_{bar}, x_{bar})$ (when considering $A<A_c$ instead). To state it differently, the barrier opens and positions $|x|>x_{bar}$ becomes consequently accessible. Start from any position belonging to the interval $I$: it can be shown that the system employs a finite time to reach the barrier at $x=x_{bar}$ (resp. $x=-x_{bar}$), when $A>0$ 
(resp. $x=-x_{bar}$ ). From here one, and following the mechanism evoked above, the system wanders within the interval $(+x_{bar}, +\infty)$ (resp. $(-\infty,-x_{bar})$). This implies that there is no overlap between the domains of definition of the two asymptotic 
probability distributions, which can be analytically estimated for each of the identified scenarios. More specifically, one can adapt to the present setting the analysis developed above, focusing on the relevant case study $+A>A_c$, to get

\begin{itemize}
\item $P_{s}(x)=\frac{1}{N}e^{-\frac{2}{{\epsilon}^{2}} U_{eff,+A}(x)}$ for $x\in\left(x_{bar},+\infty\right)$
\item $P_s(x)=0$  $\forall$ $x<x_{bar}$
\end{itemize}

The distribution $P_s(x)$ should be normalized within the interval of pertinence, $\left(x_{bar},+\infty\right)$. Here, $P_s(x)$ is obtained by integrating eq. (\ref{eq:Ueff+-A}) that was previously shown to yield eq. (\ref{eq:UeffpmASist3}). Hence, $U_{eff,+ A}(x)$ takes the same explicit form that was recovered for $A<A_c$ . The divergences, displayed by $U_{eff,+A}(x)$ at $A=A_c$, are a hallmark of the switch between the two explored dynamical regimes. Analogous conclusions apply to the dual framework with $A<0$, leading to:

\begin{itemize}
\item $P_{s}(x)=\frac{1}{N}e^{-\frac{2}{{\epsilon}^{2}} U_{eff,-A}(x)}$ for $x\in\left(-\infty,-x_{bar}\right)$
\item $P_s(x)=0$  $\forall$ $x>-x_{bar}$
\end{itemize}

Observe that the asymptotic probability to find the system inside the interval $I$ is identically equal to zero. As already remarked, the two probability distributions belong in fact to disjoint intervals. These latter intervals define the asymptotic attractors of the dynamics. In this regime, the imposed external forcing  is strong enough to break the reflection symmetry $x\rightarrow -x$. Thus, for fixed $A$, the system can never cross the barrier to eventually head backwards towards the farther edge of the (inaccessible) interval $I$. Equivalently, the required time for the system to reach the opposite barrier, diverges to infinite ($\langle\tau(- A)\rangle=+\infty$). In this regime ($A>A_c$) condition (\ref{eq:BeSuVucond}) reduces therefore to: 
\begin{equation}
\langle\tau(+ A)\rangle<<\frac{\pi}{\Omega}
\end{equation} 

\begin{figure}
\begin{subfigure}
{\includegraphics[width = 1\textwidth]{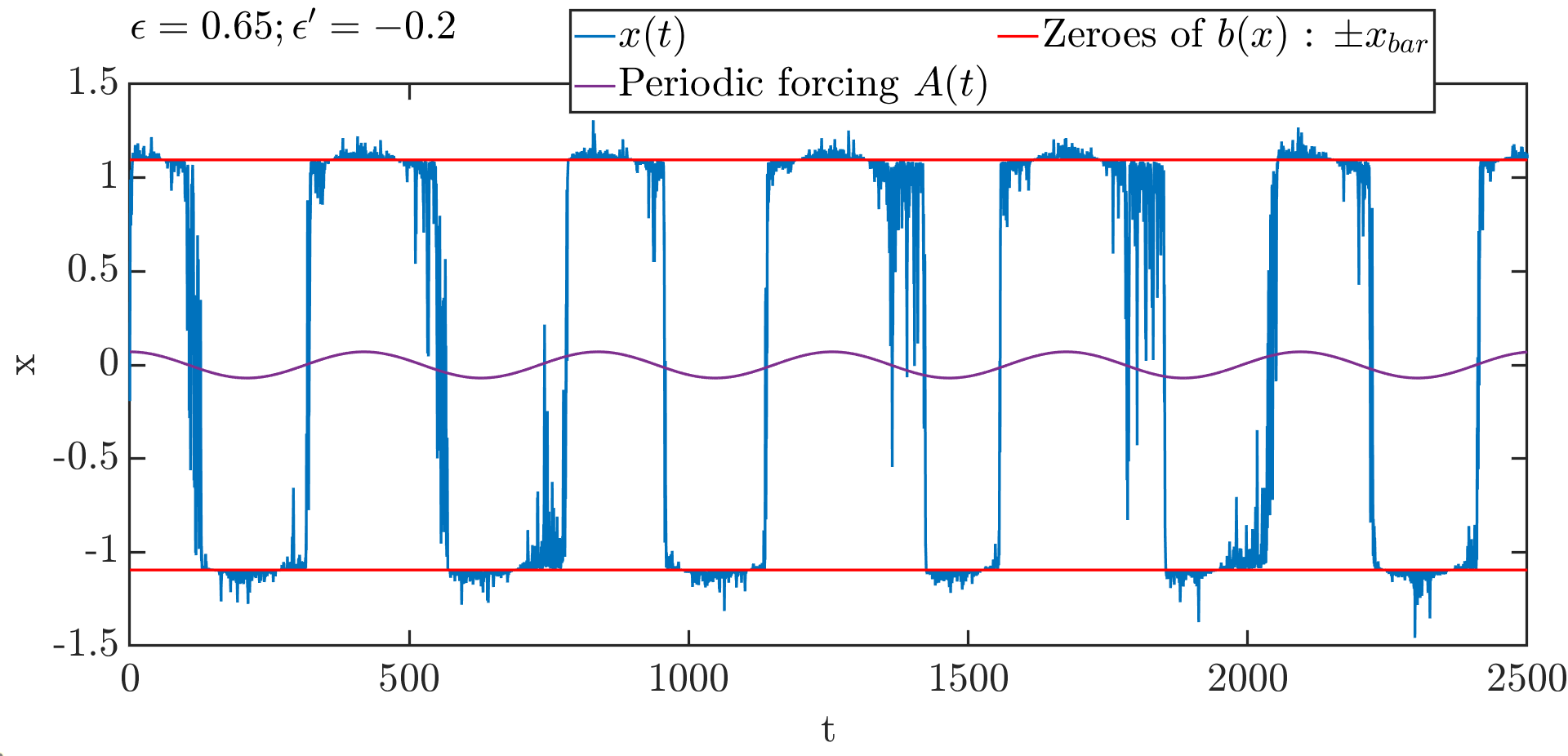}
}
\end{subfigure} \quad
\begin{subfigure}
{\includegraphics[width=1\textwidth]{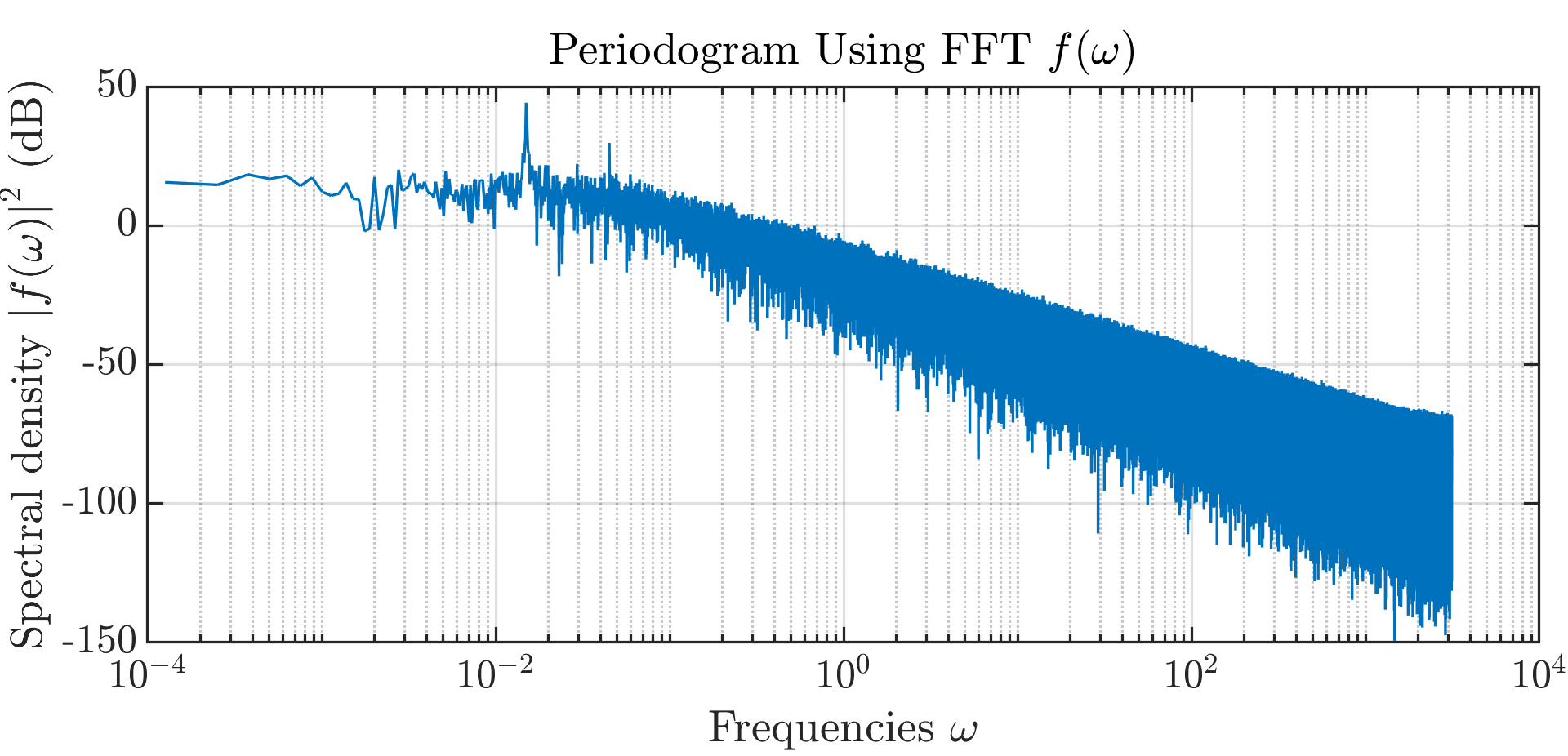}}
\end{subfigure}
\caption{Temporal series and power spectrum for $\epsilon=0.65,\epsilon'=-0.2,A=0.07> A_c\simeq 0.0438,\Omega=0.015$; $\epsilon$ and $\Omega$ are set to have the system operated in the resonant regime with the opening of the barriers. The time series displays evident periodicity. In the spectral density a peak is observed at the frequency $\omega=\Omega$.  The amplification of the input signal is quantified as $\frac{x_{max}}{A}\simeq\frac{1}{A}\simeq 14$, for this specific choice of $A$.}
\label{fig:Sys1NewStocRes}
\end{figure}

It is consequently anticipated that the latter is the sole condition to be met for  the sought resonance to develop, when the external forcing is periodically modulated in between $-A$ and $A$. In other words, the period of the external forcing should be large enough to allow the  
system to relax towards the new attractor, once the sign of the periodic forcing has been changed. Working in this setting, no lower bound on detectable frequencies is found. \\
Let us now focus again on the simplified toy-model that we have taken as a reference to illustrate the conclusion of our analysis. In Figure \ref{fig:Sys1NewStocRes} a representative time series produced for the system operated under the generalized regime $A>A_c$, is being reported. This is in particular obtained for the same choice of parameters employed in Figure \ref{fig:Sys1StocRes1}, with the notable exception of $A$. The resulting dynamics is clearly periodic, as also testified by the peak displayed by the power spectrum at $\omega = \Omega$. Remarkably, the measured amplification factor is still noteworthy. The opening of the barriers leads to the emergence of a resonant regime, which displays the characteristics of a classical stochastic resonance phenomenon, in a setting that enables us to relax the condition on the lower bound of detectable frequencies. 

We end this Section with a few remarks. Firstly, the region out of the barriers may be  inaccessible by construction for many systems. For instance, the barriers may correspond to physical boundaries which cannot be crossed by the evolving populations. Secondly, the choice of the potential $V(x)$ here operated  is rather specific and, at the same time, crucial. Indeed, the inflection points allow for the existence of finite portions of the accessible space where the deterministic dynamics is significantly slowed down. These latter domains can be made arbitrarily close to the barriers by tuning the parameter $\epsilon '$ in $b(x)$ and this is a relevant condition that should be met for the onset of the observed synchronized dynamics.

\section{Conclusions}

We have here revisited the celebrated stochastic resonance paradigm, with the aim of relaxing the conditions that are conventionally invoked for the phenomenon to materialize. In general, stochastic resonance deals with a  one dimensional bistable system, shaked by external additive noise. When perturbed by a small periodic drive, and for a suitable choice of the noise amplitude, the system oscillates regularly between the two minima of the deterministic potential. Remarkably, successive swappings are synchronous to the imposed periodic modulation. This is a resonant mechanism which can be exploited to 
unravel small periodic signals, otherwise too weak for direct detection. Starting from these premises, we have here revisisted the conventional framework by assuming a setting where bistability sets in as a byproduct of the nonlinear nature of the multiplicative noise. More into details, we assumed a system subject to a non linear potential with just one minimum. The multiplicative noise is instead engineered so as to slow down the dynamics of the system in correspondence of specific locations which can be hence depicted as effective - noise seeded - fixed points. For a proper choice of the parameters involved, the examined system jumps erratically between the two aforementioned dynamical attractors, thus displaying a genuine bistable behaviour. When subject to an external sinusoidal perturbation, the system swings regularly between the two fictitious fixed points, the pace of the swapping being set by the periodicity of the imposed forcing. This is the analogous of the classical stochastic resonance, for a system confined in symmetric and  unimodal potential well. As a further conclusion, we also identified a different resonant regime, which we eventually traced back to the opening/closing of the barriers which define the regions of trapping. Interestingly, working under this generalized perspective no lower bound exists for the frequencies which can be detected, for any given noise intensity. In future perspective, it will be interesting to recast the above conclusions in the relevant setting where the multiplicative nature of the noise arises from finite size corrections of an inherently discrete system of interacting individuals.

\section{Acknowledgements}
Work supported by NEXTGENERATIONEU (NGEU) and funded by the Ministry of University and Research (MUR), National Recovery and Resilience Plan (NRRP), project MNESYS (PE0000006) – A Multiscale integrated approach to the study of the nervous system in health and disease (DN. 1553 11.10.2022)

%\printbibliography
\bibliographystyle{ieeetr}
\bibliography{mybibliography}

\end{document}